\documentclass[prd,amsfonts,twocolumn,nofootinbib,showpacs]{revtex4}
\RequirePackage{graphicx}
\usepackage{enumerate}

\begin{document}
\title{What we can learn from the spectral index of the tensor mode}

\author{Kazunori Kohri}
\affiliation{Cosmophysics group, Theory Center, IPNS, KEK,  
and The Graduate University for Advanced Study (Sokendai), Tsukuba
305-0801, Japan}
\author{Chia-Min Lin}
\affiliation{College Entrance Examination Center (CEEC), Taipei
10673, Taiwan} 
\author{Tomohiro Matsuda}
\affiliation{Laboratory of Physics, Saitama Institute of Technology, Fukaya, Saitama 369-0293, Japan}

\begin{abstract}
If the beginning of inflation is defined at the moment when the vacuum
 energy of the inflaton starts to dominate, the energy density of the
 other fields at that moment is (by definition) comparable to the
 inflaton. 
Although the fraction will be small at the horizon
 exit due to the inflationary expansion, they can alter the scale
 dependence of the spectrum. 
At the same time, velocity of the inflaton field may not coincide with
the slow-roll (attractor) velocity.
Those dynamics could be ubiquitous but can easily alter the
scale dependence of the spectrum.
Since the scale dependence is currently used to constrain or even
 exclude inflation models, it is very important to measure its shift,
which is due to the dynamics that does not appear in the original
 inflation model.
Considering typical examples, we show that the spectral index of the tensor
 mode is a useful measure of such effect.
Precise measurement of the higher runnings of the scalar mode will be
 helpful in discriminating the source.
\end{abstract}

\maketitle

\section{Introduction}
Scale-dependence of the spectrum of the curvature perturbation has been
used to discriminate inflationary scenarios~\cite{Lyth:1998xn, Olive:1989nu}.
In addition to the scalar mode, recent discovery of the B-mode
polarization\cite{Polarbear2014, BICEP22014, Hinshaw:2012aka, Ade:2013uln}
 has ignited studies of the tensor modes.
Although the detection of the inflationary tensor mode by the BICEP2 was
not successful, it stimulated study of the mechanism of
generating peculiar scale-dependence~\cite{recent-st,
KM-ambiguity, Matsuda:2008fk}. 
Besides those recent trends in inflationary cosmology, models of
particle physics (e.g, supersymmetric models and string theory) are expecting
a large amount of scalar fields that may be dynamical in the very early
stage of inflation. 
In the light of multi-field inflation, those extra degrees of freedom
may alter the scale-dependence of both the tensor
 and the scalar perturbations.
Consequently, they may bring in a kind of ambiguity to the inflationary
parameters.
Even if one is able to assume that the inflaton is the only
dynamical field, it is still hard to believe that the inflaton velocity 
``before'' inflation coincides with the slow-roll velocity ``during''
 inflation.
Initially the inflaton velocity could deviate 
from the slow-roll (attractor) velocity and
it will lead to the shift of the scale-dependence.
These effects are very common and they can be
responsible for the scale dependence of the spectrum.

We start with the obvious situation.
The model shown in Fig.\ref{fig1} is a textbook example, which helps
understand why it is very easy to shift the scale-dependence of the
spectrum.
We focus on the simple and common dynamics.
Since the scale dependence is currently used to constrain inflation
models, it is very important to consider the shift of the scale dependence
which could be caused by the dynamics that does not appear in the original
 inflation model.
We call such dynamics the ``scalon'', since it contributes only to the
generation of the scale dependence. 
Our study can be discriminated from other attempts in which non-trivial
interactions play the key role.
We are considering ubiquitous remnants.
We also consider the case in which the field is not
negligible at the end of inflation.
The model is presented in Fig.\ref{fig2}.
Finally we consider single-field inflation in which small deviation from
the slow-roll velocity can cause significant shift of 
the scale dependence.
We show that in those cases the observation of the spectral index of the
tensor mode will play important role in finding the scalon contribution. 
More precise measurement of the higher
runnings of the scalar spectrum will be helpful for the discrimination.

\section{How to measure and remove the scalon contribution in the spectrum}
We briefly explain the overview before the details.
We are focusing on the generality of the situation and considering
simple dynamics of a scalar field.
We thus avoid introduction of specific interactions and non-trivial kinetic
terms.
This point could be distinguishable from the 
other recent attempts~\cite{recent-st}.

Single-field inflation ``usually'' expects $r+8n_t=0$, where $r$ is the
tensor-to-scalar ratio and $n_t$ is the spectral index of the tensor mode.
The relation could be violated if there are multiple
scalar fields during inflation.\footnote{Later we will show that it is
possible to violate the relation in single-field inflation.}
Intuitively we are expecting simple cases for multi-field inflation
explained below.
\begin{itemize}
\item Consider trajectory of Fig.\ref{fig1}.
      In this scenario the scalon is short-lived. The curvature
      perturbation evolves during inflation~\cite{matsuda-elliptic}. 
\item Consider trajectory of Fig.\ref{fig2}. 
      In this scenario the scalon remains dynamical until the end of inflation.
      The curvature perturbation is generated at the end.
\end{itemize}
      In both cases the curvature perturbation at the pivot scale is
      unchanged, while the scale dependence is
      shifted by the 
      scalon. 

      Before explaining more details, we point out that similar (but
      opposite) situation already appeared in the curvaton 
      model. In the simplest curvaton model, in which the slow-roll
      parameters of the curvaton field is negligible (i.e, the curvaton
      dynamics cannot generate the scale dependence), the scale dependence  
      is generated by the dynamics of the inflaton field whose
      perturbation is negligible in the curvaton model.
      In that case the inflaton dynamics contributes only to the scale
      dependence.
      Note that the field (or the dynamics) that contributes only to the
      scale dependence may commonly appear in any inflationary model.
      Similar effect may appear without adding a scalar field, as we 
      will describe the situation below for single-field inflation.

\begin{itemize}
\item Inflaton velocity may have deviation
      from the slow-roll velocity. 
      In that case the curvature perturbation is known to evolve after
      horizon exit, since the 
      additional degree of freedom (i.e, the deviation) forms the decaying
      mode. In that case $r+8n_t=0$ is not satisfied in single-field inflation.

\end{itemize}
Deviation from the slow-roll velocity may have serious impact on the scale
dependence. We will consider this possibility in Sec.\ref{dev-slow}.
The sign of the spectral index could be reversed.

The relation $r+8n_t=0$ is useless in the curvaton
      scenario, since the form of the curvature perturbation is
      completely different.
      We are not discussing those ``alternative'' models, including
      modulation (e.g. modulated reheating).

      In general, ``multi-field inflation'' includes models in which 
      multiple perturbations contribute the curvature perturbation.
      Such models are extensively studied since they may have a significant
      evidence of multi-field inflation in the relation among non-linear
      parameters~\cite{Run_in_curvaton}.
      To avoid confusions  in this paper, those models are specifically called 
      ``mixed perturbation scenario''.
      In this paper we are avoiding mixed perturbation scenarios since
      the non-Gaussianity parameter is small.
      
      One might notice that the current CMB data is sometimes used to
      distinguish multi and single-field inflation.
      The major reason could be that there is the stringent upper bound on the
      isocurvature perturbation.
      Although an additional field ``can'' generate significant
      isocurvature perturbations, it is not mandatory.
      In other cases, as we have stated above, ``multi-field inflation''
      sometimes means ``mixed perturbation scenario''.
      To avoid those confusions we emphasize that ``inflation with the
      scalon dynamics'' will not be distinguishable at this moment; we mean
      that the experimental plans at hand will hardly find the
      distinction. 
      However, since what we call ``the scalon'' is the very common dynamics
      that could appear in the early Universe, we claim that it has to
      be identified and removed by 
      more future experiments.

\subsection{``Modest'' multi-field inflation}
\label{modest-multi}
\begin{figure}[t]
\centering
\includegraphics[width=1.0\columnwidth]{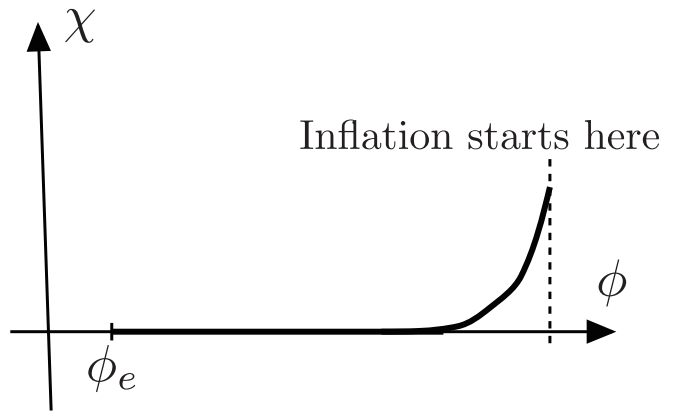}
 \caption{Schematic picture of the ``modest'' multi-field inflation
 model. The scalon soon disappears and the number of e-foldings is determined by
 $\phi$. Mixing in the spectrum is negligible.}
\label{fig1}
\end{figure}

We start with the simplest case.
We introduce the scalon field (additional free scalar field $\chi$),
which is dynamical at the beginning of inflation.
One can imagine the case in which $\dot{\chi}_*\gg\dot{\phi}_*$ at horizon
exit but $\dot{\chi}\ll \dot{\phi}$ at the end. 
Here ``$*$''  denotes the value at the horizon exit.
See also Fig.\ref{fig1}.
Using the standard definition of the curvature perturbation, the scalon
determine the initial adiabatic curvature perturbation. 
We also consider the case with $\dot{\chi}_*<\dot{\phi}_*$, which can also
shift the scale dependence.

We are considering multi-field inflation.
However, since we are considering a ``temporal'' field in this section, the derivative of the e-foldings with
respect to the extra field ($N_{,\chi}$) is negligible.\footnote{In this
section we are trying to explain the situation as intuitively as possible. 
See Ref.\cite{Byrnes:2006fr} if details are needed.}
Here the subscript with comma denotes derivative with respect to the field.
Cases with significant $N_{,\chi}$ has been considered in
Ref.\cite{Gong:2007ha}.
Remember that we are focusing on the simple dynamics that does not
assume specific interactions. 
We will assume canonical kinetic term for the fields.
We are avoiding mixing of the fields in the potential.
The scalon field is separated from the inflaton $\phi$ and ceases to
be dynamical before the end of inflation. 
In that way our analysis does not depend on specific inflationary model.
One can add $\chi$ to any (single-field, canonical
kinetic term) inflationary model on one's choice.

When a couple of fields $\phi$ and $\chi$ are dynamical during inflation,
the spectral index of the curvature perturbation is calculated as~\cite{Byrnes:2006fr}
\begin{eqnarray}
\label{original-index}
n_s-1&\simeq& -(6-4\cos^2 K_A)\epsilon_H +2\eta_{\sigma\sigma} \sin^2
 K_A\nonumber\\
&& 
+4\eta_{\sigma s}\sin K_A\cos K_A 
 +2\eta_{ss}\cos^2 K_A,
\end{eqnarray}
where the subscripts $\sigma$ and $s$ denote the adiabatic and entropy
directions at the horizon exit.
We choose $\dot{\sigma}^2=\dot{\chi}^2+\dot{\phi}^2$.
Here the definitions of the slow-roll parameters are
\begin{eqnarray}
\label{slow-def}
\epsilon_a &\equiv& \frac{M_p^2}{2}\left(\frac{V_{,a}}{3H^2
				       M_p^2}\right)_*^2
= \frac{1}{2M_p^2}\left(\frac{V_{,a}}{3H^2}\right)_*^2\nonumber\\
\eta_a&\equiv& \left[\frac{V_{,aa}}{3H^2}\right]_*\nonumber\\
\eta_{ab}&\equiv& \left[\frac{V_{,ab}}{3H^2}\right]_*,
\end{eqnarray}
where $H$ is the Hubble parameter, $M_p$ is the reduced Planck mass
and $a, b, ..$ denote $\phi, \chi, \sigma$ and $s$.
We introduce separation given by $\epsilon_H\equiv
-[\dot{H}/H^2]_*=\epsilon_\phi+\epsilon_\chi$. 
 $K_A$ is defined using the transfer matrix and is given by
\begin{equation}
\cos K_A =\frac{T_{RS}}{\sqrt{1+T^2_{RS}}},
\end{equation}
where $T_{RS}$ is an element of the transfer matrix, which describes the
evolution of the adiabatic and the isocurvature perturbations.
We define the instantaneous adiabatic and entropy perturbations as
\begin{eqnarray}
\delta \sigma&=& \delta \phi\cos \theta  +\delta \chi \sin\theta 
\\ 
\delta s&=& -\delta \phi \sin \theta  +\delta \chi\cos\theta ,
\end{eqnarray}
where $\tan\theta =\dot{\chi}/\dot{\phi}$.
Also we define the curvature and entropy perturbations
\begin{eqnarray}
R&=& H\frac{\delta \sigma}{\dot{\sigma}}\\
S&=&  H\frac{\delta s}{\dot{\sigma}}.
\end{eqnarray}
The curvature perturbation after horizon exit is expressed as
\begin{equation}
R=[R_*+T_{RS}S_*].
\end{equation}
Here $R$ is the observable.
Since here we are assuming ``temporal'' $\dot{\chi}_*\ne 0$ at a moment, 
adiabatic velocity soon reaches $\dot{\sigma}\simeq \dot{\phi}$ after
horizon exit.
The final curvature perturbation recovers the original single-field
inflation model~\cite{Byrnes:2006fr}: $R\simeq
\left[\frac{H}{\dot{\phi}}\delta \phi\right]_*$. 

Let us see more details about the scalon contributions.
If $\dot{\chi}_*\gg \dot{\phi}_*$, the trajectory has a turn during
inflation.
Initially the adiabatic direction is $\dot{\sigma}_*\simeq\dot{\chi}_*$
but it will be $\dot{\sigma}\simeq \dot{\phi}$ before the end of inflation.
Then the curvature perturbation becomes
$R\simeq\left[\frac{H}{\dot{\phi}}\delta \phi\right]_*\gg R_*\simeq
\left[\frac{H}{\dot{\chi}}\delta \chi\right]_*$, which is possible since $T_{RS}\gg 1$.
Then one will find $\cos K_A \sim 1$ and the spectral index given
by
\begin{equation}
\label{eq-multiindex-bend}
n_s-1\simeq -2\epsilon_H +2\eta_\phi.
\end{equation}

The opposite limit ($\dot{\chi}_*\ll \dot{\phi}_*$) is the standard
single-field inflation.
One will find $\sin K_A \sim 1$ and an almost straight trajectory.
The spectral index will be 
\begin{equation}
\label{eq-multiindex-straight}
n_s-1\simeq -6\epsilon_H +2\eta_\phi.
\end{equation}

Seeing the above discrepancy between (\ref{eq-multiindex-bend}) and
(\ref{eq-multiindex-straight}), one may find that the situation is
intuitively similar to the curvaton.
More radical situation can be found in the inflating curvaton~\cite{Infcurv}.
In the above formalism one has to take into account
the scale-dependence of the transfer function $T_{RS}$.

The latter case ($\dot{\chi}_*\ll \dot{\phi}_*$
already at the beginning) seems to be the same as the conventional
single-field scenario.
However, as far as $\dot{\chi}_*\ne 0$, there is the
contribution caused by $\epsilon_\chi\ne 0$.
Although in the spectral index the contribution of the scalon could be
negligible, the shift in the higher runnings can be significant since
$\dot{\epsilon}_\chi\gg \dot{\epsilon}_\phi$ is possible~\cite{KM-ambiguity}.
We will be back to this topic in Sec.\ref{running}.

Below, we will examine the relation $r+8n_t=0$ when $\epsilon_\chi\ne 0$.
Since one extra degree of freedom is added to the original 
scenario, we need another observable (i.e, an independent equation) that
can fix the ambiguity.
The spectral index of the tensor mode can play the role.

The spectrum of the tensor perturbation is $P_{\cal T}^{1/2}=H/(2\pi)$,
which gives the tensor to scalar ratio 
\begin{equation}
r=\frac{P_{\cal T}}{P_{\cal R}}.
\end{equation}
When $\dot{\chi}_* \simeq 0$, we find conventional result 
$r=16\epsilon_{\phi}$ and $n_r\equiv \frac{d \log r}{d\log
k}=4\epsilon_H-2\eta_\phi$.
However, when $\dot{\chi}_*\gg\dot{\phi}_*$, things are not so trivial.
To check the consistency of the calculation it is very useful to
consider the equations related to $n_r$.
Therefore, in this paragraph we introduce $n_r$ in addition to $n_t$.
Remember that the running of $r$ (i.e, $n_r$) can ``directly'' be
evaluated using 
indices of the scalar and the tensor modes as\footnote{For our purpose
we are omitting details. Ref.\cite{Gong:2014qga, Gao:2014fva} will be helpful.}
\begin{equation}
\label{eq-index-r}
n_r\equiv
\frac{d \log r}{d\log k}=1-n_s+n_t, 
\end{equation}
where $n_t=-2\epsilon_H$.
We use this relation to check the consistency of the calculation below.
Let us see more details of the relations between parameters, and see how
one can remove the ambiguity.
\begin{itemize}
\item We first consider $\dot{\chi}_*\ll \dot{\phi}_*$.
Using $d\ln k =Hdt$ and the definitions of the slow-roll parameters
      (\ref{slow-def}),  one will find
\begin{eqnarray}
\frac{1}{d\ln k}\epsilon_{\phi}
&=&4\epsilon_\phi\epsilon_H-2\eta_\phi\epsilon_\phi,
\end{eqnarray}
which leads to $n_r= 4\epsilon_H -2\eta_\phi$.
Of course the result is consistent with Eq.(\ref{eq-index-r}).
Then, from the spectral index (\ref{eq-multiindex-straight}) and
      $n_t=-2\epsilon_H$, 
      one can evaluate the slow-roll parameters from 
      the observables as
\begin{eqnarray}
\epsilon_H&=&-\frac{n_t}{2}\nonumber\\
\epsilon_\phi&=&\frac{r}{16}\nonumber\\
\eta_\phi&=&\frac{(n_s-1)-3n_t}{2}\nonumber\\
\epsilon_\chi&=&-\frac{n_t}{2}-\frac{r}{16}.
\end{eqnarray}
Here $r+8n_t\ne 0$ will be the sign of $\epsilon_H\ne \epsilon_\phi$.

\item  Consider the case with $\dot{\chi}_*\gg\dot{\phi}_*$, in
       which $\chi$ soon ceases to be dynamical during inflation.
In this case we are expecting $\epsilon_H\sim \epsilon_\chi$ (i.e, the
       scalon is dominating $\epsilon_H$).
From the direct calculation (\ref{eq-index-r}), one will find $n_r= -2\eta_\phi$.
Although rather complicated, one can evaluate the same result by
considering $T_{RS}$~\cite{Byrnes:2006fr}.
To distinguish the scale dependence, it would be useful to
       define $r\equiv 16\epsilon_{\phi**}$ and evaluate 
\begin{eqnarray}
\epsilon_H&=&-\frac{n_t}{2}\nonumber\\
\epsilon_{\phi**}&=&\frac{r}{16}\nonumber\\
\eta_\phi&=&-\frac{n_r}{2}=\frac{(n_s-1)-n_t}{2}\nonumber\\
\epsilon_H-\epsilon_{\phi**}&=&-\frac{n_t}{2}-\frac{r}{16}.
\end{eqnarray}
Again, discrepancy $\epsilon_H-\epsilon_{\phi**}\ne 0$ is the
sign of an extra dynamical field.
\end{itemize}
Note that in both limits $r+8n_t\ne 0$ is the sign of the additional
       dynamical field.
Although $\chi$ introduces ambiguity to the cosmological parameters, 
the contribution can be removed if $n_t$ is measurable.

The above scenario is a well-known example of multi-field inflation.
Although in the past studies we could not find explicit argument that
relates the spectral index of 
the tensor mode to the ambiguity in the inflationary parameters, 
the above results should have been known among scientists in this field.
Below, we will consider less simple examples for which the shift of the
scale dependence has not been discussed in detail before.

\subsection{Hybrid inflation and other models}
\label{hybrid-sec}
The usual multi-field model~\cite{Byrnes:2006fr} considers
``evolution'' of the curvature perturbation during inflation.
In the model represented in Fig.\ref{fig1}, this effect compensates
the curvature perturbation. 
The situation we are going to consider in this section is rather different from
such scenario.
We introduce the scalon that can survive until the end of inflation.
Then, the initial curvature perturbation is compensated by the
curvature perturbation generated at the end.

Typically the scenario of generating the curvature perturbation at the
end of inflation is considered for hybrid-type potential, since the
original Lyth's model~\cite{Lyth:2005qk} considers modulation of the
waterfall ($\delta \phi_e\ne 0$).
In that way the Lyth's scenario requires interaction with the waterfall field.
However, our scenario {\bf does not require interaction} and works with many
variety of inflationary models in 
which the potential is given by $V=V_0+V(\phi)$ ($V_0\gg V(\phi)$).
This includes Higgs inflation.
Although the result is quite conceivable in the light of the $\delta N$
formalism, it is not obvious how the initial curvature
perturbation can be compensated at the end.
Therefore, to avoid confusions we first review the discussion in
Ref.\cite{Matsuda:2012kk}.
The calculation applies to other non-hybrid models.

First consider the inflaton $\phi$ and the waterfall field
$\sigma_w$ with the hybrid-type potential given by
\begin{equation}
\label{base-pot}
V(\phi,\sigma_w)=\frac{\lambda^2}{4}\left(\sigma_w^2-M^2\right)^2
+\frac{g^2}{2}\sigma_w^2 \phi^2 +\frac{1}{2}m_\phi^2 \phi^2.
\end{equation}
Suppose that inflation starts with $\phi>\phi_c$ and the waterfall
begins at $\phi=\phi_c$.
The critical point $\phi_c$ is $\phi_c\equiv \frac{\lambda}{g}M$.
The number of e-foldings is given by
\begin{equation}
N=H\int_{\phi_*}^{\phi_e}\frac{d\phi}{\dot{\phi}}\simeq\frac{1}{\eta_\phi}\ln
\frac{\phi_*}{\phi_c}.
\end{equation}
Remember that {\bf the original Lyth's model~\cite{Lyth:2005qk} assumes 
additional field $\chi$ that couples to the waterfall field}.
Namely, if one introduces $\chi$ that has the
same interaction as $\phi$, the end of inflation 
is defined as 
\begin{equation}
\phi^2+\chi^2=\left(\frac{\lambda M}{g}\right)^2.
\end{equation}
Then, if $\chi$ is lighter than the inflaton, the entropy
perturbation $\delta s\simeq \delta \chi\ne 0$ creates the
perturbation $\delta \phi = -\frac{\chi}{\phi_c}\delta \chi$ at
the end.
Consequently, the perturbation of the number of e-foldings created at
the end of inflation is given by
\begin{equation}
\delta N_e \equiv -\left[\frac{1}{\eta}\frac{\delta \phi}{\phi}
\right]_e
\ne 0.
\end{equation}
This is the usual scenario of ``generating the curvature perturbation at
the end''.
\begin{figure}[t]
\centering
\includegraphics[width=1.0\columnwidth]{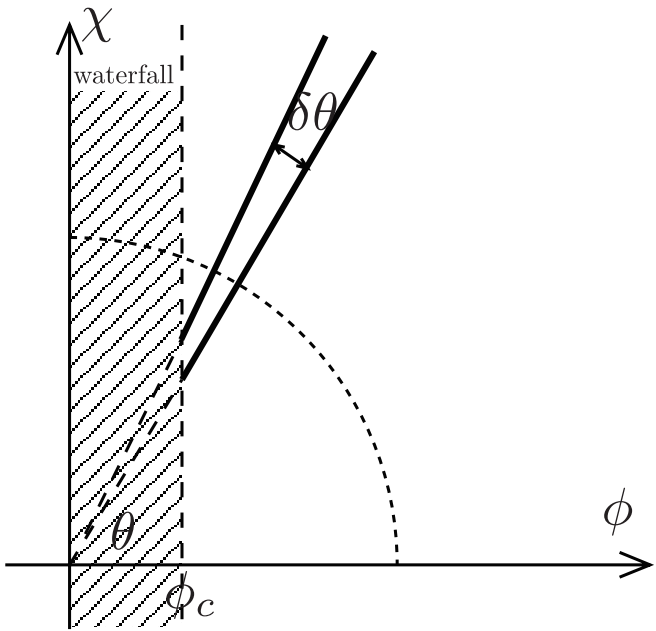}
 \caption{Hybrid inflation with a ubiquitous (non-interacting) scalar
 field. Entropy perturbation ($\delta \theta$) causes $\delta N$ at the end.}
\label{fig2}
\end{figure}

{\em In contrast to the original scenario, we are introducing a ubiquitous
field, which is decoupled from the waterfall field.} Therefore, we are {\bf
not} expecting $\delta \phi_c\ne 0$.
Just for the simplest example, we consider 
\begin{equation}
V(\phi,\sigma_w)=\frac{\lambda^2}{4}\left(\sigma_w^2-M^2\right)^2
+\frac{g^2}{2}\sigma_w^2 \phi^2 +\frac{1}{2}m^2 (\phi^2+\chi^2).
\end{equation}
Note that {\bf unlike} the usual multi-field extension of the
hybrid-type potential we are 
omitting interaction $\sim \sigma_w^2\chi^2$.
Degeneracy of the mass term ($m_\chi=m_\phi\equiv m$) makes the
trajectory straight 
and avoids the effect considered in
Sec.\ref{modest-multi}.
The adiabatic field is defined as $\sigma^2 \equiv \phi^2 + \chi^2$,
which gives 
\begin{eqnarray}
\phi&=&\sigma \cos \theta\\
\chi &=& \sigma \sin\theta.
\end{eqnarray}
We find the end {\bf for the adiabatic field} is 
\begin{equation}
\sigma_e(\theta)\equiv \frac{\phi_c}{\cos\theta}=\frac{\lambda M}{g\cos\theta},
\end{equation}
which is perturbed when $\delta \theta\simeq \delta s/\sigma \ne 0$.
(Note that $\delta \theta$ does not evolve during inflation when
$m_\chi=m_\phi\equiv m$.)
Therefore, although $\phi_c$ is not perturbed, the entropy
perturbation ($\delta \theta \ne 0$) 
causes $\delta N_e\ne 0$ at the end.
The curvature perturbation generated at the end of inflation is thus given by
\begin{eqnarray}
\delta N_e&\equiv& H\frac{\delta \sigma_e}{\dot{\sigma}_e}
\simeq
\frac{1}{\eta}\tan\theta
\left(\frac{\delta s}{\sigma}\right)_*.
\end{eqnarray}
Considering perturbation generated at the horizon exit:
\begin{equation}
\delta N_* = \left[\frac{1}{\eta} \frac{\delta \sigma}{\sigma}\right]_*,
\end{equation}
we find the ratio between ``initial'' and ``at the end'': 
\begin{eqnarray}
\left|\frac{\delta N_e}{\delta N_*}\right|
\simeq \tan\theta.
\end{eqnarray}
Here $|\delta s_*|\simeq|\delta \sigma_*|$ is used for the calculation.

More intuitive argument is possible.
Considering similar triangles in Fig.\ref{fig2}, we find the
number of e-foldings 
\begin{equation}
N=\frac{1}{\eta}\ln
\frac{\sigma_*}{\sigma_e}=\frac{1}{\eta}\ln
\frac{\phi_*}{\phi_c}.
\end{equation}
Therefore, intuitively the $\delta N$ formalism suggests
\begin{eqnarray}
\delta N&=&\frac{1}{\eta}\frac{\delta \phi_*}{\phi_*}\nonumber\\
&=&\frac{1}{\eta}\frac{\delta \sigma_*}{\sigma_*}
-\frac{1}{\eta}\frac{\delta \sigma_e}{\sigma_e}.
\end{eqnarray}

Let us see what happens if $\theta \sim \pi/2$. 
Then the final curvature perturbation is dominated by $\delta N_e$,
which reproduces the ``conventional'' perturbation 
$\delta N\simeq \frac{1}{\eta}\frac{\delta \phi}{\phi}$.
Here, we already know that the scalon can shift the  scale-dependence of
the spectrum.
Separating the slow-roll parameter
$\epsilon_H=\epsilon_\phi+\epsilon_{\chi}$,
 we find the spectral index given by~\cite{Lyth:2005qk}
\begin{equation}
n_s-1\simeq -2\epsilon_H +2\eta_\phi.
\end{equation}
In the opposite case, when $\theta \sim 0$, generation of the curvature perturbation
at the end is negligible and we find $n_s=-6\epsilon_H+2\eta$.
In both cases $\epsilon_H$ is shifted by $\epsilon_\chi$.
The result is quite similar to the conventional ``multi-field'' model
in Fig.\ref{fig1},
although the mechanism of generating the curvature perturbation is utterly different. 
The property of the scalon is also different.
For the model considered in this section, $\dot{\chi}$ can be
significant until the end of inflation.

Usually the typical hybrid-type model is excluded because its spectrum
is blue ($n_s-1 \sim 2\eta_\phi>0$).
However, as we have seen above, one can introduce a ubiquitous field
$\chi$ to change the spectral index into red ($n_s-1<0$)~\cite{Matsuda:2012kk}.
Again, the spectral index of the tensor mode is needed to distinguish the scalon
contribution.

The situation presented in Fig.\ref{fig2} is quite general.
The mechanism can be applied to any inflationary model in which $\chi$
causes negligible change in $\phi_c$.
Imagine a potential like $V=V_0+V(\phi)$ with $V_0\gg V(\phi)$, which is
very common.

\subsection{Deviation from the slow-roll (single-field inflation)}
\label{dev-slow}
Introduction of the ``extra degree of freedom that is responsible for
the scale dependence of the curvature perturbation'' does not always
require additional scalar field. 
It may appear as a parameter measuring deviation from the inflationary attractor.
First, remember that at the very beginning of inflation the inflaton
velocity ($\dot{\phi}$) may not coincide with the slow-roll velocity ($\dot{\phi}_s$). 
Moreover, if the inflaton is moving fast in the opposite direction
before the onset of inflation, inflaton may even stop during
inflation~\cite{Seto:1999jc, Bassett:2005xm, Takamizu:2010je}. 
Although the analysis could be slightly model-dependent, the essential
of the argument is quite simple.
If the inflaton velocity deviates from the
slow-roll velocity defined there (i.e, $\dot{\phi}_* \ne \dot{\phi}_s(t_*)\equiv
-V_{,\phi}(\phi_*)/3H$), the curvature perturbation converges to the value
evaluated using the slow-roll velocity.
Namely, one will find the evolution of the curvature perturbation
$R_*=\frac{H}{\dot{\phi}_*}\delta \phi\rightarrow 
R=\frac{H}{\dot{\phi}_s}\delta \phi$ before the end of inflation.
Intuitively the above result is quite conceivable in the light of the $\delta
N$ formalism.

Since the time-dependence of the quantity $\phi$ is not
determined by $\dot{\phi}_s$, the scale-dependence (i.e, evolution of
$\phi$ and $H$ that determines $k$-dependence of the curvature
perturbation)
 will be different from the standard scenario.
Unlike the multi-field model, the
correction in the spectral index will appear not only from
 $\dot{H}$ (i.e, $\epsilon_H$) but also from $\dot{\phi}\ne \dot{\phi}_s$.
Namely, if one uses $d\ln k =Hdt$ for the calculation, there will be corrections like 
\begin{equation}
\dot{\epsilon}_\phi=\dot{\phi}_*\frac{d\epsilon_\phi}{d\phi}\equiv
\left[1+R_{D*}\right]\dot{\epsilon}_\phi^{(0)},
\end{equation}
where we defined $R_D\equiv \frac{\dot{\phi}-\dot{\phi}_s}{\dot{\phi}_s}$, and
$\dot{\epsilon}^{(0)}_\phi\equiv \dot{\phi}_s\frac{d\epsilon_\phi}{d\phi}$.
Note that ``$^{(0)}$'' is used for quantities without deviation, except
for the Hubble parameter that always includes $\dot{\phi}\ne
\dot{\phi}_s$ for the kinetic energy.
See Eq.(\ref{slow-def}) for the definitions of the slow-roll parameters. 
Our definition is $\epsilon_\phi^{(0)}\equiv
\frac{1}{2M_p^2}\left(\frac{V_{,\phi}}{3H^2}\right)_*^2$. 
In the same way, we also find 
\begin{equation}
\label{eHplusdelta}
\epsilon_H=(1+R_{D*})\epsilon_\phi^{(0)}+\Delta
\epsilon_H,
\end{equation}
where $\Delta \epsilon_H$ comes from the time derivative of
the kinetic energy in the Hubble parameter.
Since the kinetic energy of the inflaton is not identical to the
attractor solution, $\ddot{\phi}$ could not be negligible.
Correction to the spectral index is 
\begin{equation}
n_s-1 = (1+R_D)(n_s^{(0)}-1) -6\Delta \epsilon_H.
\end{equation}
If the Hubble parameter can be written as $3M_p^2
H^2=\frac{1}{2}\dot{\phi^2}+V(\phi)$, we find
\begin{eqnarray}
\Delta \epsilon_H&=&-\frac{\dot{\phi}_*\ddot{\phi}_*}{6M_p^2H^3}\nonumber\\
&\simeq&\epsilon_\phi^{(0)}R_{D*}(1+R_{D*}),
\end{eqnarray}
where we used $\dot{\phi}=\dot{\phi}_s(R_D+1)$,
$\ddot{\phi}=\dot{\phi}_s\dot{R}_D$,
$\dot{R}_D\simeq -3HR_D$ and $\epsilon_\phi^{(0)}=\frac{\dot{\phi}^2_s}{2H^2M_p^2}$.
Here we expect $\frac{d}{dt}\left(\dot{\phi}-\dot{\phi}_s\right)\sim -3H
\left(\dot{\phi}-\dot{\phi}_s\right)$ for the deviation.

If the inflaton ``stops'' during inflation one will find $R_D\sim
-1$, which gives $n_s\simeq 0$.
The result is consistent with the naive intuition:
scale dependence disappears when inflaton stops.

Since $R_D$ changes quickly during inflation, there will be 
a signature of $R_D\ne 0$ in the running of the spectral index.
One will find 
\begin{eqnarray}
\alpha_s&\equiv& \frac{d n_s}{d\ln k}\simeq (1+R_{D*})\alpha_s^{(0)}
-3R_{D*} (n_s^{(0)}-1)\nonumber\\
&&-6R_{D*}\left[-3\epsilon_\phi^{(0)}
+4\epsilon_\phi^{(0)}\epsilon_H-2\eta_\phi^{(0)}\epsilon_\phi^{(0)}\right]\nonumber\\ 
&&-6R_{D*}^2\left[-6\epsilon_\phi^{(0)}+4\epsilon_\phi^{(0)}\epsilon_H-2\eta_\phi^{(0)}\epsilon_\phi^{(0)}
\right].
\end{eqnarray}
Contribution related to $R_{D*}$ mainly
appears from $\sim -3R_{D*} (n_s^{(0)}-1)+18R_{D*}\epsilon_\phi^{(0)}
+36R_{D*}^2\epsilon_\phi^{(0)}$.
Since $\alpha_s>0$ is almost excluded by the observations, we can see
that $R_{D*}>0$ is not a realistic scenario.

Note that the blue spectrum of hybrid-type inflation ($n_s-1\simeq
2\eta_\phi >0$ for $\epsilon_\phi\simeq 0$)
can be turned into red when $R_{D*}<-1$.
Therefore the ambiguity related to the deviation is quite serious.
Even though $r$ in hybrid inflation could be very small and the tensor
mode could not be seen in the observations, 
more precise measurement of $\alpha_s$ can be used to exclude such
possibility. 

$r+8n_t=0$ is violated because $\epsilon_H$ is not
identical to $\epsilon_\phi^{(0)}$.
Here $r=16\epsilon_\phi^{(0)}$ and $n_t=-2\epsilon_H$.
Note that according to the discussion in Ref.\cite{Seto:1999jc} one will
find that $r=16\epsilon_H$ is not correct here.
Again, the additional parameter $R_{D*}$ can be fixed if $n_t$ could be
observed.
Since Eq.(\ref{eHplusdelta}) can be rewritten as
$\epsilon_H\simeq(1+R_{D*})^2\epsilon_\phi^{(0)}$, we find
$(1+R_{D*})^2=-\frac{8n_t}{r}$.

\subsection{Higher runnings of the modestly fast-rolling $\chi$ and the
  problem of the initial condition}
\label{running}
The extra dynamical field can change the scale dependence of the spectrum.
The situation could be serious if there are many scalar fields, whose
effective masses are $O(H)$ or less during inflation.
In this section we estimate the possible contributions from such
remnants.
Here  we assume
$\eta_\chi\lesssim{\cal O}(0.1)$.
In that case the 
conventional slow-roll approximations are still valid for the estimation.
If one considers the quadratic potential $V(\chi)=\frac{1}{2}m_\chi^2
\chi^2$ for the fast-rolling field $\chi$, one will
find~\cite{Dimopoulos:2003ce}   
\begin{eqnarray}
\chi(t)&=& \chi_0e^{-Kt},
\end{eqnarray}
where 
\begin{eqnarray}
K&\equiv&
 \frac{3}{2}H\left[1-\sqrt{1-\frac{4}{9}\left(\frac{m_\chi^2}{H^2}\right)} 
\right].
\end{eqnarray}
This gives
\begin{eqnarray}
\dot{\rho}_\chi&\simeq&-K\left(-\frac{K^2}{m_\chi^2}+1\right)\rho_\chi,
\end{eqnarray}
where contribution from the kinetic energy has not been neglected.
Just for simplicity we consider $\dot{\rho}_A\simeq-K\rho_A$.

Note that $\chi$ can change the spectral index and
predict $\Delta \alpha_s\ne 0$ at the same time.
We find
\begin{eqnarray}
\Delta \alpha_s&\simeq&-6\epsilon_\chi\left(4\epsilon_H-2\eta_\chi\right),
\end{eqnarray}
which could be negative if $2\epsilon_H>\eta_\chi$
or simply $\eta_\chi<0$~\cite{KM-ambiguity}.

\section{Conclusion and discussion}
\label{conclusion}
If the Universe started with a chaotic state, there could be many
dynamical fields before the onset of inflation.
Inflationary expansion starts when the vacuum energy of the inflaton starts to
dominate.
At that moment the other dynamical fields may have energy density comparable
to the inflaton vacuum energy.
If one wants to disregard their effect, these fields must be inflated
away before the onset of the last 60 e-foldings.
This is not a trivial assumption since their effect could not be negligible
even if their fraction is small.\footnote{When the field value of $\chi$ is larger than the Planck scale, the
      fraction of $\rho_\chi$ is $\sim \frac{m_\chi^2
      \chi^2/2}{3H^2 M_p^2}\sim
      \eta_\chi\left(\frac{\chi}{M_p}\right)^2$. 
      This cannot be a ``small fraction'' when the slow-roll parameters of
      $\chi$ are not negligible. We are not assuming such significant
      modification 
      of the original inflationary model.}

In this paper we first considered a non-interacting
field $\chi$ and calculated the possible shift of the scale dependence.
The first model is presented in Fig.\ref{fig1}, which we
considered as a textbook example.
In this case the additional inflaton field ($\chi$) may disappear during
inflation but it can significantly change the scale dependence of the
spectrum. 
The second model is presented in Fig.\ref{fig2}, which is less known but 
is showing a very common situation.
In both cases our result illuminates the importance of the observation
of the spectral index of the tensor mode.
We also considered a single-field inflation model in which deviation from the
slow-roll is not negligible.
We found a similar effect and showed that observation of the tensor mode
is important.
Note that in the last scenario the shift of the scale-dependence is
considered in a single-field inflation model.

In this paper we have chosen specific scalon dynamics that are very
common in the early Universe, and showed that the ambiguity caused by
the remnant is crucial for the scientific argument.
Although our results are encouraging and seemingly suggesting the next
step, unfortunately in reality the tensor mode of the  
inflationary perturbations has not been
observed yet and the present experimental plans are not designed to
determine the spectral index of the tensor mode accurately enough to
distinguish the source of the scale dependence.
Nevertheless, we hopefully expect that more future observations will be
able to distinguish the scalon effect and our study will be useful for
identifying the inflation model.

\section*{Acknowledgement}
K.K. is supported in part by Grant-in- Aid for Scientific
research from the Ministry of Education, Science, Sports, and
Culture (MEXT), Japan, Nos. 23540327 ,26105520, 26247042, 15H05889(K.K.)
The work of K.K. is also supported by the Center for the Promotion of
Integrated Science (CPIS) of Sokendai (1HB5804100).

\end{document}